\documentclass[aps,prx,twocolumn,amsmath,amssymb]{revtex4}
\usepackage{graphicx}
\usepackage{Jonasmacros}
\usepackage{mathptmx}
\usepackage{txfonts}
\usepackage{bm,bbold}
 
\usepackage[colorlinks,citecolor=red,linkcolor=Fuchsia]{hyperref}
\usepackage[usenames,dvipsnames,svgnames]{xcolor}

\newcommand{\tobedeleted}[1]{\textcolor{green}{#1}}
\renewcommand{\tobedeleted}[1]{\relax}

\begin{document}
\date{\today}
\title{Universal voltage scaling due to self-averaging of the quantum corrections in graphene}

\author{R. Somphonsane}
\email{ratchanok.so@kmitl.ac.th}
\affiliation{Department of Physics, King Mongkut's Institute of Technology Ladkrabang, Bangkok 10520, Thailand}

\author{H. Ramamoorthy}
\author{G. He}
\author{J. Nathawat}
\author{S. Yin}
\affiliation{Department of Electrical Engineering, University at Buffalo, the State University of New York, Buffalo, NY 14260-1900, USA}
\author{J. P. Bird}
\affiliation{Department of Electrical Engineering, University at Buffalo, the State University of New York, Buffalo, NY 14260-1900, USA}

\author{C.-P. Kwan}
\author{N. Arabchigavkani}
\author{B. Barut}
\affiliation{Department of Physics, University at Buffalo, the State University of New York, Buffalo, NY 14260-1500, USA}

\author{M. Zhao}
\author{Z. Jin}
\affiliation{High-Frequency High-Voltage Device and Integrated Circuits Center, Institute of Microelectronics of Chinese Academy of Sciences, 3 Beitucheng West Road, Chaoyang District, Beijing, PR China}

\author{J. Fransson}
\email{jonas.fransson@physics.uu.se}
\affiliation{Department of Physics and Astronomy, Uppsala University, Box 516, SE-751 21 Uppsala, Sweden}

\begin{abstract}
The differential conductance of graphene is shown to exhibit a zero-bias anomaly at low temperatures, arising from a suppression of the quantum corrections due to weak localization and electron interactions. A simple rescaling of these data, free of any adjustable parameters, shows that this anomaly exhibits a universal, temperature- ($T$) independent form. According to this, the differential conductance is approximately constant at small voltages ($V<k_BT/e$), while at larger voltages it increases logarithmically with the applied bias, reflecting a quenching of the quantum corrections. For theoretical insight into the origins of this behavior, we formulate a model for weak-localization in the presence of nonlinear transport. According to this, the voltage applied under nonequilibrium induces unavoidable dephasing, arising from a self-averaging of the diffusing electron waves responsible for transport. By establishing the manner in which the quantum corrections are suppressed in graphene, our study will be of broad relevance to the investigation of nonequilibrium transport in mesoscopic systems in general. This includes systems implemented from conventional metals and semiconductors, as well as those realized using other two-dimensional semiconductors and topological insulators.  
\end{abstract}
\maketitle


\section{Introduction}
\label{sec-introduction}
It has long been understood that the conductance of mesoscopic systems may exhibit quantum corrections at low temperatures, arising from the combined influence of weak localization \cite{PhysRevB.28.2914,PhysRep.107.1,RevModPhys.57.287} and electron interactions \cite{RevModPhys.57.287,PhysRevLett.44.1288,PhysRevB.35.4205}. The former phenomenon \cite{PhysRevB.28.2914} is due to the coherent interference of time-reversed pairs of Feynman paths, which return to their origin after a sequence of elastic scattering events, thereby enhancing the resistance above its Drude value. The interaction correction \cite{PhysRep.107.1}, on the other hand, has been discussed in terms of the scattering generated by the \emph{charge hologram} associated with such closed paths \cite{PhysRevB.35.4205}. While these corrections have long been the subject of study in normal metals and semiconductors, interest in these phenomena has been revived in recent years due to their manifestations in emergent two-dimensional materials, with the most notable example being provided by graphene. The unusual aspects of the bandstructure of this material, including its linear energy dispersion and the chiral nature of its carriers, significantly modify weak localization, in a manner that has been discussed in a number of theoretical \cite{PhysRevLett.97.146805,PhysRevLett.97.036802,PhysRevLett.97.196804,PhysRevLett.98.176806,PhysRevLett.101.126801,SolStComm.151.1550,PhysRevLett.108.166606} and experimental \cite{PhysRevLett.97.016801,PhysRevLett.98.136801,PhysRevLett.98.176805,PhysRevLett.100.056802,PhysRevB.78.125409,ApplPhysLett.93.122102,PhysRevLett.103.226801,NewJPhys.11.095021,Nanotechnol.21.274014,JPCM.22.205301,NatMater.10.443,ApplPhysLett.103.143111} works. Most notable here is that the details of the localization are strongly dependent upon the nature of the impurities in the system, with exact backscattering being forbidden for remote impurities that generate long-range scattering \cite{JPhysSocJpn.67.2857,JPhysSocJpn.74.777}. As such, this behavior corresponds to weak antilocalization, a phenomenon that is normally associated with materials with strong spin-orbit coupling \cite{PhysRevB.28.2914,PhysRep.107.1,RevModPhys.57.287}. The antilocalization occurs in spite of the very weak spin-orbit coupling in graphene, but is suppressed in the presence of short-ranged impurities; these restore weak localization by allowing backscattering between the inequivalent $K$ and $K'$ valleys. Elsewhere, other works have explored the nature of the interaction-related correction to graphene's conductance \cite{PhysRevB.82.075424,PhysRevB.83.195417,NewJPhys.13.113005,PhysRevLett.108.106601,PhysRevB.88.235406,PhysRevB.90.035423}, and have found it to be of similar magnitude to that arising from weak localization.

In both the original experimental work on quantum corrections in normal metals and semiconductors \cite{PhysRevB.28.2914,PhysRep.107.1,RevModPhys.57.287,PhysRevB.35.4205}, and in more recent investigations performed on graphene \cite{PhysRevLett.97.016801,PhysRevLett.98.136801,PhysRevLett.98.176805,PhysRevLett.100.056802,PhysRevB.78.125409,ApplPhysLett.93.122102,PhysRevLett.103.226801,NewJPhys.11.095021,Nanotechnol.21.274014,JPCM.22.205301,NatMater.10.443,ApplPhysLett.103.143111,PhysRevB.82.075424,PhysRevB.83.195417,NewJPhys.13.113005,PhysRevLett.108.106601,PhysRevB.88.235406,PhysRevB.90.035423}, the primary emphasis has been on obtaining information on these phenomena by studying their influence on the \emph{linear} conductance. More specifically, most works have addressed the manner in which the corrections are affected by a magnetic field, which breaks time-reversal symmetry and suppresses weak localization while leaving the interaction contribution unaffected \cite{PhysRevB.28.2914,PhysRep.107.1,RevModPhys.57.287,PhysRevLett.44.1288,PhysRevB.35.4205}. In contrast, far fewer studies have explored the manner in which these phenomena are affected under  nonequilibrium conditions. (Notable exceptions include early works that demonstrated the inability of an electric field to break time reversal during coherent backscattering \cite{PhysRevLett.43.721,PhysRevB.25.5563}, and later experiments on the differential conductance of GaAs/AlGaAs quantum dots \cite{PhysRevB.55.4061} and short metallic nanobridges \cite{PhysRevLett.70.841,PhysRevB.63.165426}.) While there have been a few investigations of the nonlinear differential conductance of graphene \cite{PhysRevB.82.045411,PhysRevB.83.205421,PhysRevB.84.245427,PhysRevB.85.161411,PhysRevLett.109.056805,SciRep.3.3533}, there is still relatively little that is understood about the manner in which the quantum corrections in this material (and in other Dirac materials) are affected under nonequilibrium conditions. It is this specific problem that we address here, from both experimental and theoretical perspectives.

The experimental component of this work involves studies of the differential conductance ($g$) of graphene transistors, implemented in both monolayer and bilayer material. At low temperatures ($T$), where quantum corrections are expected to influence transport, the zero-bias conductance ($G$) of these devices is suppressed by the combined influence of weak localization and electron interactions. Application of a nonzero voltage ($V$) quenches these phenomena, however, and leads to an enhancement of the differential conductance that defines a zero-bias anomaly. By implementing a simple rescaling of these data, in which we plot the bias-induced change of differential conductance ($\Delta g = g(V)-G$) as a function of the dimensionless voltage ($eV/k_BT$, where $k_B$ is the Boltzmann constant), we show that the zero-bias anomaly collapses onto a universal, temperature-independent form. According to this, the linear conductance remains unchanged for voltages $eV\lesssim k_BT$, while at larger voltages it increases as a logarithmic function of $V$, reflecting the quenching of the quantum corrections. This universal voltage scaling of the quantum corrections is observed in both monolayer and bilayer devices, and on the electron and hole sides of the Dirac point. Quantitative insight into the origins of this behavior is provided in the theoretical component of this work, in which we develop a formal description of the weak-localization correction under strongly nonequilibrium conditions. This is achieved by making use of a nonequilibrium Green function approach, in which we address the influence of disorder-induced scattering up to the level of the maximally-crossed diagrams responsible for weak localization. Our essential finding is that the applied voltage introduces an additional dephasing in transport, that is analogous to that known to arise from a magnetic field or from nonzero temperature \cite{JPCM.14.R501}. More specifically, by opening an energy window for transport, the applied voltage causes a self-averaging of the electron wavefunction, according to which diffusing waves gradually decohere with one another as they propagate around the same scattering loop. While our calculations are performed for the weak-localization correction alone, the strong similarities that they exhibit with the results of our experiment suggest that the interaction correction should be similarly affected by the self-averaging phenomenon.

The organization of the remainder of this paper is as follows. Section \ref{sec-expmethods} provides a description of the graphene devices used in this study, and of the different techniques that are used to measure them. Our main experimental results are then presented in Sec. \ref{sec-expresults}, while in Sec. \ref{sec-theory} we develop a theoretical model for weak localization in graphene under nonequilibrium. In Sec. \ref{sec-discussion} we discuss some of the implications of our results, before concluding in Sec. \ref{sec-conclusions}.

\section{Experimental methods}
\label{sec-expmethods}
Graphene devices were fabricated by exfoliating Kish graphite onto a heavily-doped Si substrate with a 300-nm SiO$_2$ cap layer \cite{NanoLett.13.4305,SciRep.7.10317}. Layer identification was achieved through a combination of optical microscopy and Raman imaging \cite{SciRep.7.10317}, following which individual graphene flakes were contacted with Cr/Au (3-/50-nm) electrodes, defined by electron-beam lithography and lift-off. The conductive Si substrate served as the (back-) gate of these devices, which was biased at an appropriate voltage ($V_g$) to vary the carrier concentrations. A number of devices were fabricated, and characterized electrically, and exhibited similar and consistent characteristics. In this work we focus on a detailed study of the differential conductance of representative devices realized from monolayer and bilayer graphene. Measured electron ($\mu_e$) and hole ($\mu_h$) mobilities in these devices ($\mu_e = 12,000$ cm$^2$/Vs and $\mu_h = 14,000$ cm$^2$/Vs in monolayer, and $\mu_e = 1,300$ cm$^2$/Vs and $\mu_h = 1,100$ cm$^2$/Vs in bilayer, both at 4 K and a concentration of 10$^{12}$ cm$^{-2}$) were consistent with prior reports \cite{RevModPhys.81.109} for graphene on SiO$_2$. For four-probe measurements of the linear conductance ($G$), a small AC voltage ($v_d$) was applied between the source and drain contacts of the device, allowing the conductance to be determined from the measured AC current ($i_d$) and from the voltage ($v$) drop across a pair of internal voltage probes. An example of this configuration is provided in the left inset of Fig. \ref{Fig1} (a), which features an optical micrograph of one of our bilayer devices. The value of $v_d$ was configured to yield an internal voltage $v\sim100$ $\mu$V and the measurement frequency was set at 13 Hz. For measurements of the differential conductance, the AC conductance was determined in the same manner as above, but now with an additional DC voltage ($V_d$, applied to the source and drain contacts) superimposed upon the AC component. In all of the data presented here, the variation of differential conductance is indicated as a function of the DC voltage drop ($V$) across the pair of internal probes. This voltage was determined from the measured AC voltages, according to ($V = V_d \cdot(v/v_d)$). All measurements were made with the samples mounted in vacuum, on the cold finger of a closed-cycle cryostat.

\section{Experimental results}
\label{sec-expresults}
\begin{figure}[t]
\begin{center}
\includegraphics[width=0.99\columnwidth]{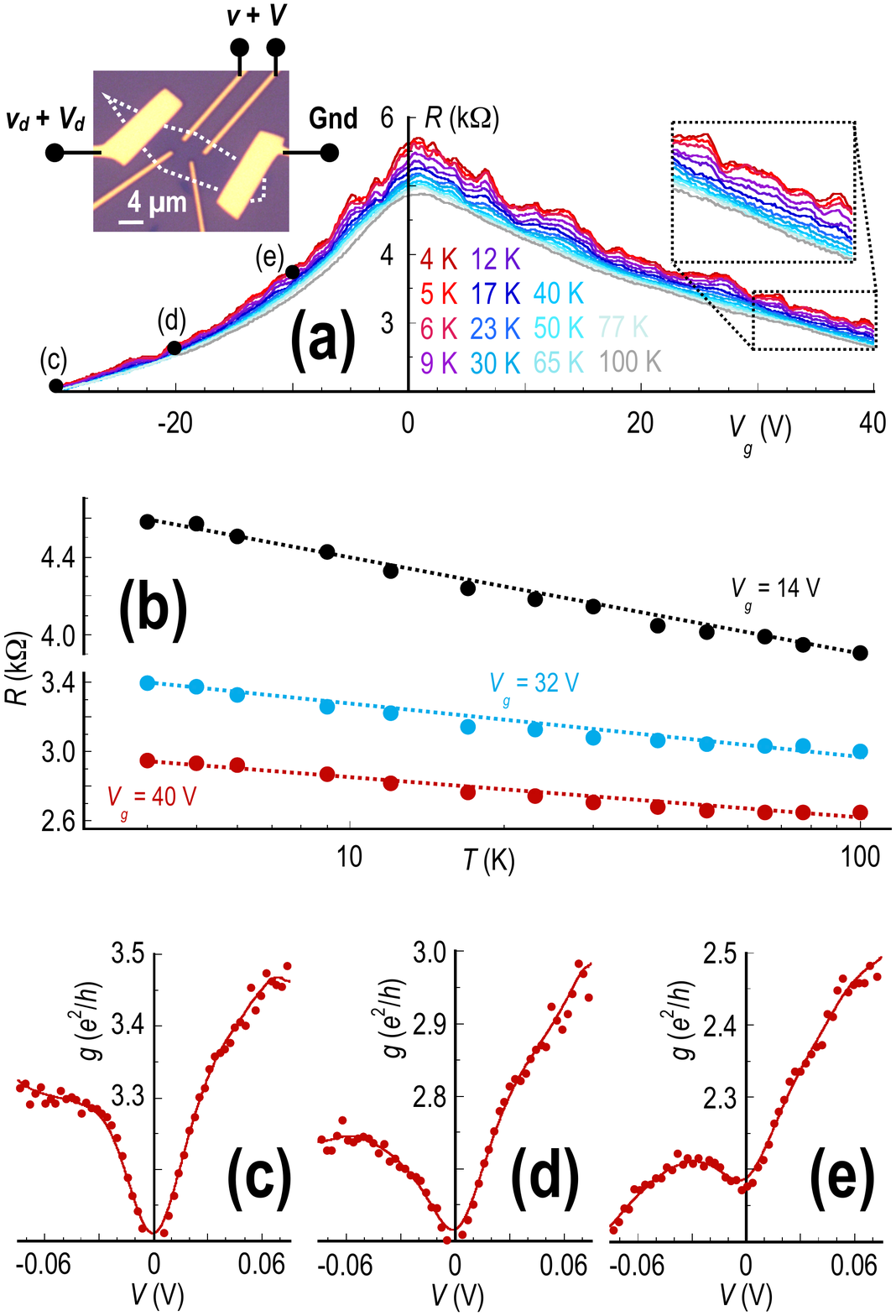}
\end{center}
\caption{(Color online) (a) The main panel shows the variation of the small-signal (AC) resistance as a function of gate voltage, for the bilayer device shown in the optical image in the left inset. This inset also indicates the biasing scheme used to measure differential conductance in a four-probe configuration. The right inset is an expanded view of the data, indicating the emergence of conductance fluctuations and the quantum corrections at low temperatures. (b) Variation of resistance as a function of temperature, determined from the results of panel (a) at three different gate voltages (indicated). The dotted lines are guides to the eye that indicate the logarithmic scaling characteristic of quantum corrections. Note the break in the vertical axis of the figure. (c) -- (e) Measurements of the differential conductance of the bilayer device at 4 K and for the three gate voltages identified in the main panel (a). Filled symbols are experimental data while the solid lines represent weighted fits through these points.
}
\label{Fig1}
\end{figure}

We begin our discussions by focusing on the manner in which the quantum corrections are manifested in the linear-transport characteristics of the devices. In the main panel of Fig. \ref{Fig1} (a), we show the variation of the small-signal resistance ($R\equiv 1/G$) of the bilayer device as a function of its gate voltage. Measurements are shown for various temperatures from 4 $-$ 100 K, and reveal several important characteristics. Firstly, as the temperature is lowered below 30 K, reproducible conductance fluctuations develop in the curves, signaling the emergence of coherent mesoscopic transport \cite{ApplPhysLett.101.093110,PhysRevB.86.161405}. More importantly, the fluctuations are superimposed upon a background trend for increasing resistance with decreasing temperature (see the right inset to Fig. \ref{Fig1} (a)), behavior that is typical of the quantum corrections \cite{PhysRevB.28.2914,PhysRep.107.1,RevModPhys.57.287,PhysRevLett.44.1288,PhysRevB.35.4205}. This connection is further established in Fig. \ref{Fig1} (b), where we plot the variation of resistance as a function of temperature at three representative gate voltages. In all three cases, the resistance varies as a logarithmic function of temperature, a characteristic signature of quantum corrections in two-dimensional materials \cite{PhysRevB.28.2914,PhysRep.107.1,RevModPhys.57.287,PhysRevLett.44.1288,PhysRevB.35.4205}.

The main emphasis in this report is on understanding the manner in which the quantum corrections in graphene are suppressed under nonequilibrium conditions. The essential phenomenon in which we will be interested is presented in Figs. \ref{Fig1} (c) -- (e), in which we plot the low-temperature (4 K) differential conductance of a bilayer device at three gate voltages (identified in the main panel of Fig. \ref{Fig1} (a)). Common to these data is a zero-bias anomaly, according to which the conductance is suppressed at zero bias but then increases when a voltage of either polarity is applied. (The anomaly is superimposed upon a broader conductance variation, which has previously been discussed as a signature of electron heating \cite{PhysRevLett.109.056805}.) As we now describe, the zero-bias anomaly is attributed to a nonequilibrium suppression of the quantum corrections in these devices.

\begin{figure}[t]
\begin{center}
\includegraphics[width=0.99\columnwidth]{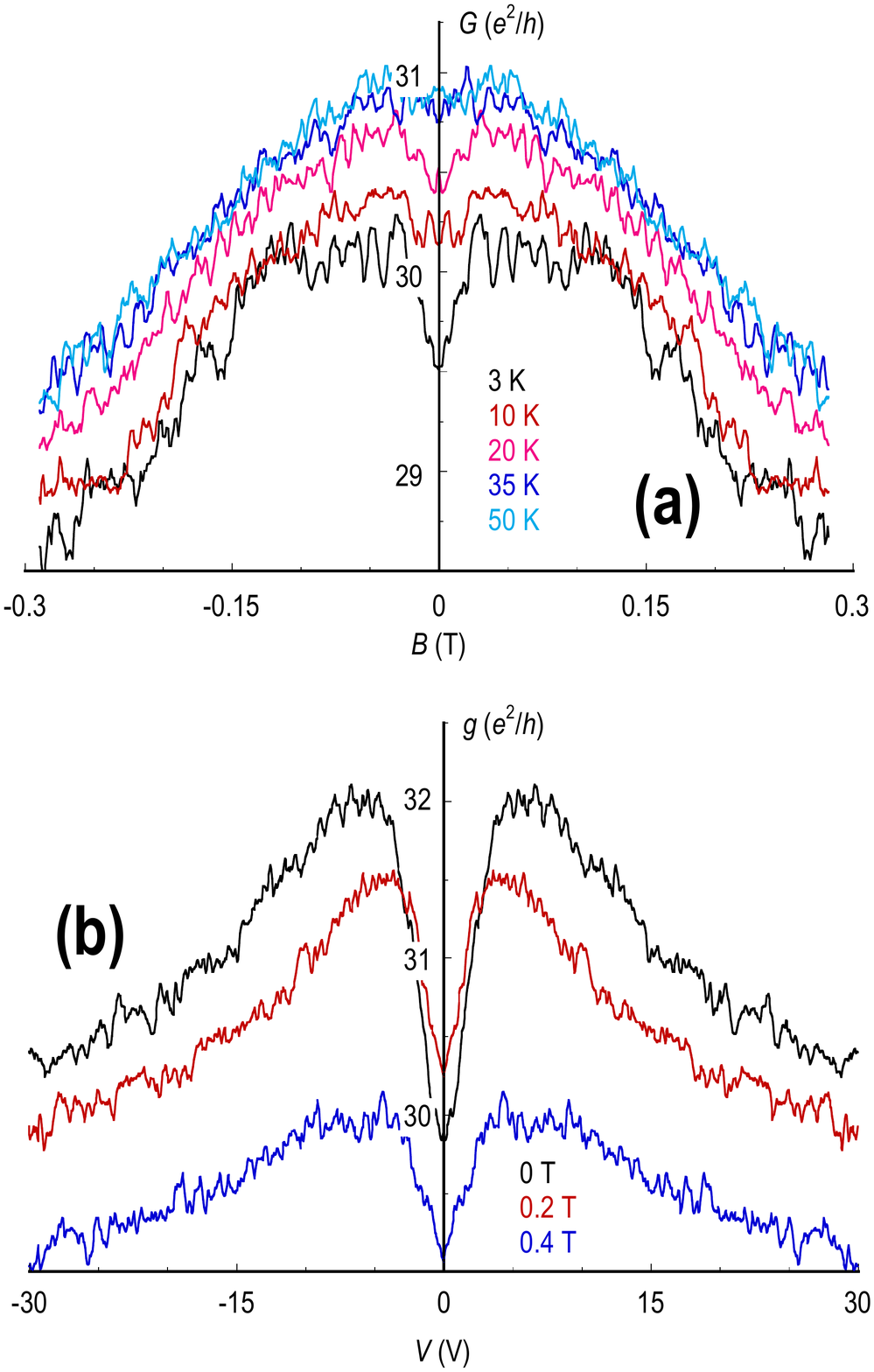}
\end{center}
\caption{(Color online) (a) Symmetric component \cite{SciRep.7.10317} of the linear magneto-conductance of the monolayer device at five different temperatures (indicated). (b) Differential conductance of the monolayer device at 3 K and at three magnetic fields. Data in panels (a) and (b) were obtained for a gate voltage at the Dirac point of the device ($V_g = 14$ V).
}
\label{Fig2}
\end{figure}

In normal investigations of quantum corrections, it is common to distinguish the influence of weak localization and electron interactions by applying a perpendicular magnetic field ($B$); as noted already, this suppresses weak localization while leaving the interaction correction unaffected \cite{PhysRevB.28.2914,PhysRep.107.1,RevModPhys.57.287,PhysRevLett.44.1288,PhysRevB.35.4205}. In Fig. \ref{Fig2} (a), we show the results of measurements of the symmetric component \cite{SciRep.7.10317} of the linear magneto-conductance of the monolayer device at five different temperatures. At 3 K, the signature of weak localization can be clearly seen in the data, in the form of a narrow conductance dip that is centered at zero magnetic field, and which coexists with reproducible fluctuations that extend over the entire field range. Both of these features weaken as the temperature is increased to 35 K, consistent with the expected reduction in carrier phase coherence \cite{JPCM.14.R501,ApplPhysLett.101.093110,PhysRevB.86.161405}. The most important point here, however, is that the localization is quenched at very weak magnetic fields, as little as 30 mT. Noting this, in Fig. \ref{Fig2} (b) we plot the results of measurements of the low-temperature differential conductance of the monolayer device at three different magnetic fields. These data were obtained for a gate voltage close to the Dirac point and at $B = 0$ T the zero-bias anomaly is close to $2e^2/h$ in size. Its amplitude is reduced under the application of the magnetic field, decreasing to around 50 \% of its original value at 0.4 T. The crucial point here is that this field value is significantly larger than that required in Fig. \ref{Fig2} (a) to suppress weak localization. Consequently, we may conclude that the zero-bias anomaly in the differential conductance arises, at zero magnetic field at least, from the combined influence of weak localization and electron interactions. Moreover, the data of Fig. \ref{Fig2} (b) suggest that the relative magnitude of these two quantum corrections should be roughly the same.

\begin{figure}[t]
\begin{center}
\includegraphics[width=0.99\columnwidth]{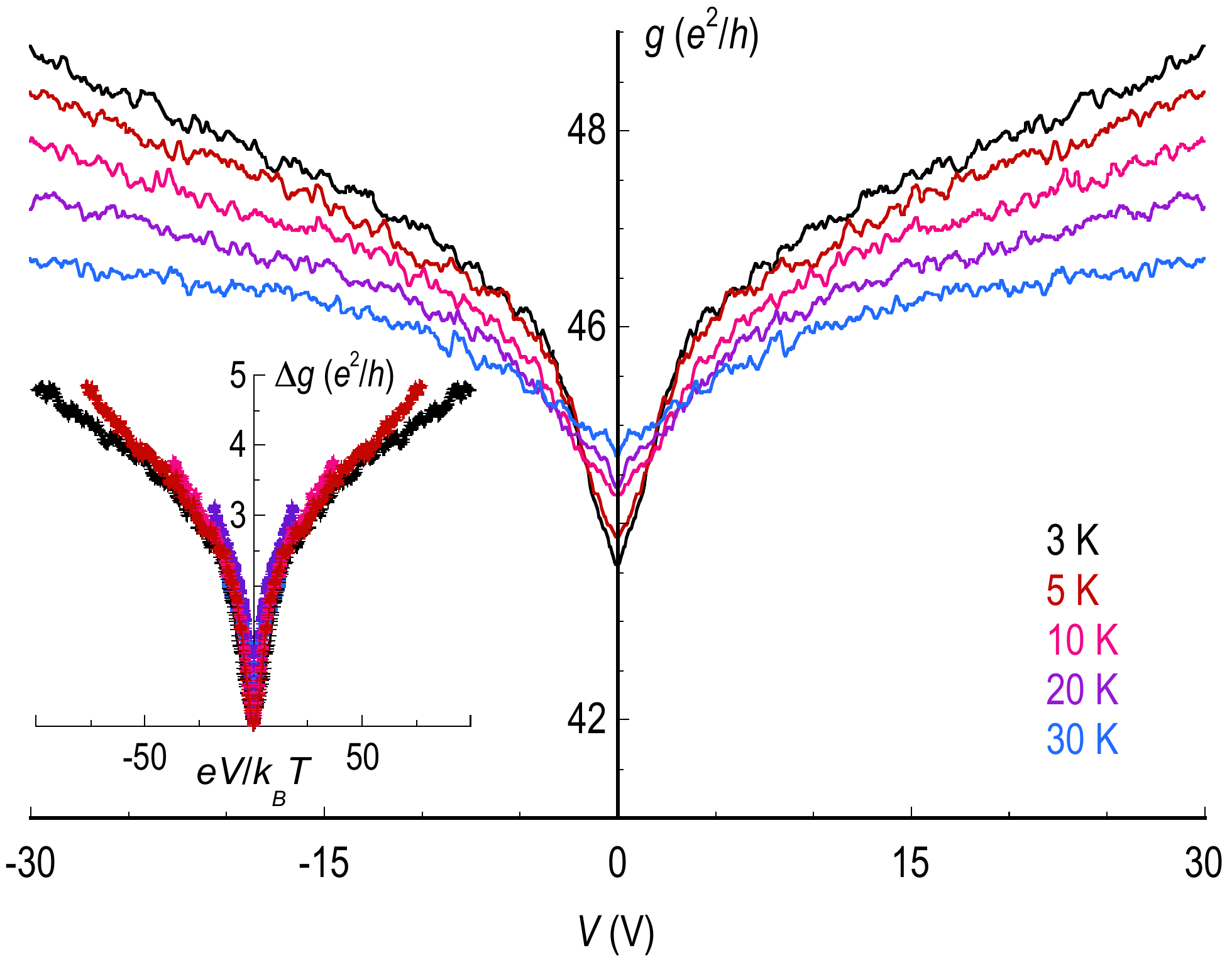}
\end{center}
\caption{The main panel shows the results of measurements of the differential conductance of the monolayer device at various temperatures from 3 $-$ 30 K. The data were obtained for a gate voltage $V_g = +6$ V, while the Dirac point in this device was positioned at $V_g = +14$ V. This condition therefore corresponds to a gate-induced hole concentration of $p = 5.8\cdot10^{11}$ cm$^{-2}$. The inset shows a rescaling of the data from the main panel, according to which we plot the variation of the bias-induced conductance change ($\Delta g = g(V)-G$) as a function of the dimensionless voltage ($eV/k_BT$). The colors of the different data points correspond to the same temperatures as indicated in the main panel.}
\label{Fig3}
\end{figure}

While the data of Figs. \ref{Fig2} (a), (b), were obtained for a monolayer device, with the gate voltage configured at the Dirac point, the general features of the zero-bias anomaly revealed in these figures were found also in our studies of bilayer graphene (as confirmed already in Fig. \ref{Fig1}), and were largely unchanged as the gate voltage was used to move the Fermi level between the valence and conduction bands.

\begin{figure}[t]
\begin{center}
\includegraphics[width=0.99\columnwidth]{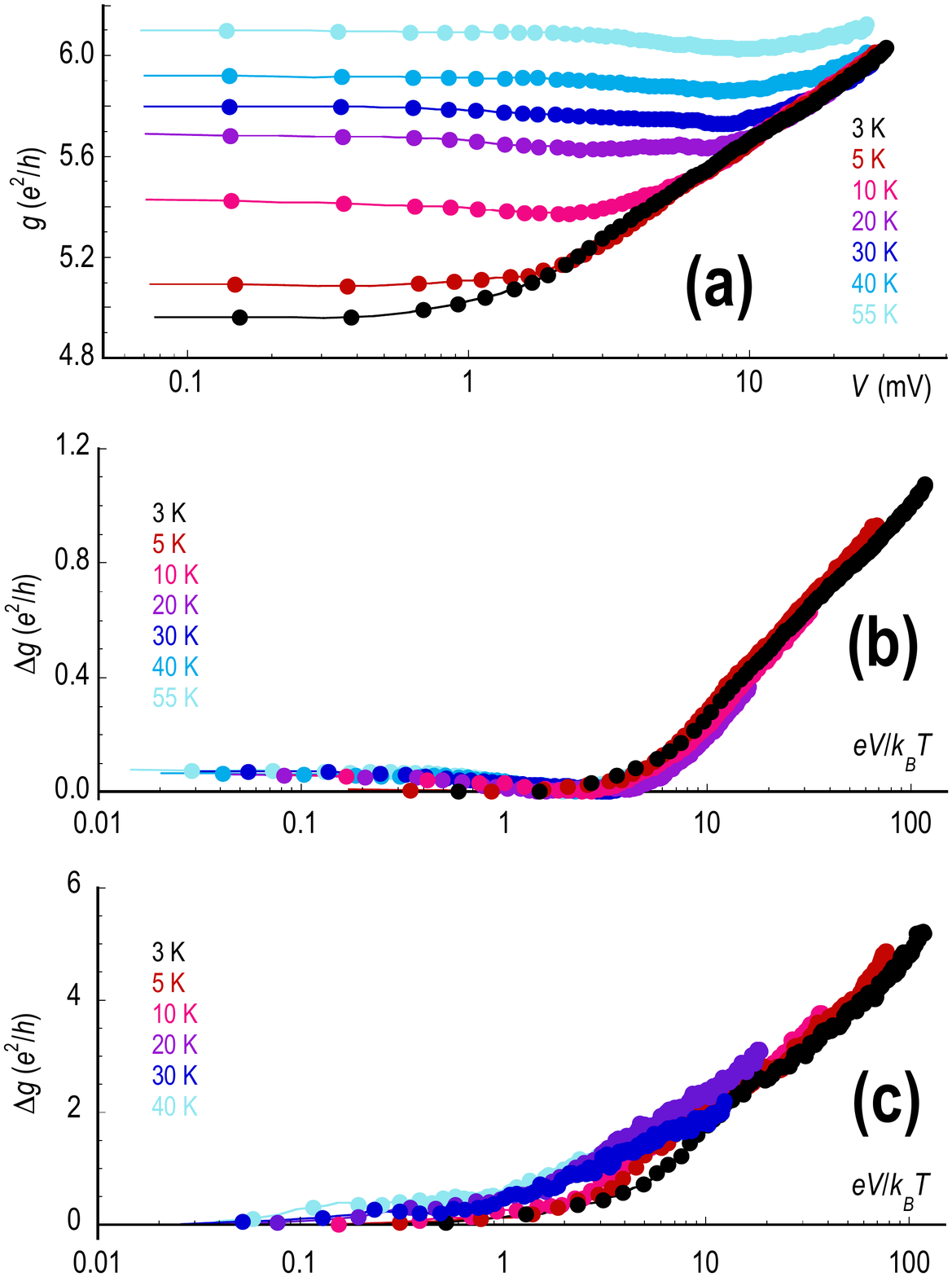}
\end{center}
\caption{Temperature dependent differential conductance of the bilayer device, measured at various temperatures from 3 $-$ 55 K. The data were obtained for a gate voltage $V_g = -24$ V, while the Dirac point in this device was positioned at $V_g = -5$ V. This condition therefore corresponds to a gate-induced hole concentration of $p = 1.4\cdot10^{12}$ cm$^{-2}$. (b) Data of panel (a), obtained after rescaling. (c) Data of Fig. \ref{Fig3}, obtained after similar rescaling.}
\label{Fig4}
\end{figure}

Consistent with the temperature-dependent nature of linear transport, as evidenced in Figs. \ref{Fig1}, \ref{Fig2}, the zero-bias anomaly in the differential conductance is also found to depend strongly upon temperature. This is illustrated in Fig. \ref{Fig3} (a), in the main panel of which we show measurements of the differential conductance of the monolayer device at a number of different temperatures. As the temperature is increased from an initial value of 3 K, two important trends are apparent in this data: firstly, the zero-bias conductance increases with increasing temperature, consistent with the expected suppression of the quantum corrections, and; secondly, the overall amplitude of the zero-bias anomaly is simultaneously reduced. Remarkably, we find that these data can be rescaled onto a common curve by implementing the procedure utilized in the inset to Fig. \ref{Fig3}. Here we replot the data of the main panel to show the variation of the bias-induced conductance change ($\Delta g = g(V)-G$) as a function of the dimensionless voltage ($eV/k_BT$). Simply by means of this parameter-free rescaling, we see that the conductance curves obtained at different temperatures essentially collapse onto one another.

For further insight into the rescaling of the differential conductance, we refer to the results of Fig. \ref{Fig4}. In contrast to the results of Fig. \ref{Fig3}, which were obtained for a monolayer device, in Figs. \ref{Fig4} (a), (b), we show the results of this rescaling for a bilayer transistor. In Fig. \ref{Fig4} (a), we show the variation of the differential conductance ($g(V)$) as a function of the applied voltage, with the abscissa indicated on a logarithmic scale. From these data alone, it is clear that the differential conductance exhibits two distinct regimes of voltage-dependent behavior; the first at low voltages where it is approximately constant, and the second at higher voltages where it instead exhibits a logarithmic variation. In Fig. \ref{Fig4} (b), we rescale these data as in the inset to Fig. \ref{Fig3}, in this case representing the dimensionless voltage on a logarithmic scale. As with the behavior revealed in Fig. \ref{Fig3}, we once again see how the curves obtained at various temperatures in Fig. \ref{Fig4} (a) collapse onto a common curve as a result of the data rescaling. Having replotted the data in this way, it is moreover apparent that the crossover between the voltage-independent and voltage-dependent regimes occurs for $eV/k_BT\sim1$, a natural result if the influence of the voltage is to open a thermally-resolved energy window for transport (a point that we return to below). In Fig. \ref{Fig4} (c), we provide yet another illustration of this rescaling, in this case for the monolayer data of Fig. \ref{Fig3}. The similarity of these results to those of Fig. \ref{Fig4} (b) is striking, with a crossover near $eV/k_BT\sim1$ and a logarithmic scaling at higher voltages.

Having established the universal voltage scaling of the quantum corrections in graphene, in the next section we develop a theoretical model to account for this behavior. Our approach involves formulating a description of the weak-localization correction under nonequilibrium, and demonstrating that this exhibits the essential characteristics of our experimental data. While a full treatment of this problem should also involve calculating the interaction-induced correction under nonequilibrium, this latter task is considered to be beyond the scope of the current work. By clarifying how weak localization is influenced under nonequilibrium, however, we gain important insight into the relevant processes responsible for the quenching of the quantum corrections in experiment.

\section{Theoretical treatment of weak localization in graphene under nonequilibrium}
\label{sec-theory}
In usual discussions of weak localization it is common to describe this phenomenon in terms of the two-particle propagator known as the Cooperon \cite{PhysRep.107.1,RevModPhys.57.287,QuantumTheorySolids,PhysRep.140.193}. This comprises a quasiparticle that follows a diffusive path, and which interferes with its time-reversed partner, providing the natural viewpoint from which to discuss weak localization. As a linear-response construct, however, the Cooperon may only be utilized under equilibrium, or quasiequilibrium, conditions, and thus is not applicable in the nonlinear regime of interest here. We therefore adopt an alternative approach to treat weak localization \cite{SolStComm.86.195,QuantumKinetics}, which is based on the use of nonequilibrium Green functions. Under spatial averaging, we obtain \emph{rainbow} diagrams that provide the impurity-limited scattering lifetime in the self-consistent Born approximation, and \emph{maximally-crossed} diagrams that account for weak localization (see App. \ref{sec-elstruct}). To implement our calculations, we consider a graphene sheet contacted by a pair of metallic leads, and calculate its electronic structure subject to the boundary conditions imposed by the coupling to these leads. To consider the details of transport in the presence of disorder, we introduce short-range scattering centers that represent atomically-sharp defects. By including such impurities in our model we are able to capture effects arising from inter-valley scattering \cite{JPhysSocJpn.67.2421,PhysRevLett.89.266603}, a key mechanism that is necessary to give rise to weak localization \cite{PhysRevLett.97.146805}. (The role of long-range disorder, such as that generated by substrate impurities, is not considered here. In this sense, our analysis is pertinent to intrinsic graphene, in the presence of local imperfections and defects, but free of any substrate interactions.)

While the details of our  calculation of the nonequilibrium weak-localization correction are provided in the appendix to this paper, the essential features of our theory are as follows. Using a tight-binding model of the monolayer graphene lattice, with imperfections arising from a random collection of short-ranged impurities, and tunneling to a pair of metallic leads, we calculate the current flowing through this layer under the application of fixed voltage. The nonequilibrium nature of this problem is taken into account by formulating the model on the Keldysh contour (see appendix for details). The reduction of current ($\delta I$) caused by weak-localization is then determined by applying an impurity averaging, in which the contribution to the current from the maximally-crossed diagrams is calculated. Using the resulting expression, the localization-induced correction to the differential conductance is ultimately expressed (in the limit $T\rightarrow0$) as:
\begin{align}
\frac{d}{dV}\delta I=&
	-\frac{2e^2}{h}
	\frac{\Gamma^L\Gamma^R}{D_c^2}
	\ln\frac{D_c}{eV}
	.
\label{eq-diddv}
\end{align}
Here, $\Gamma^{L,R}$ represent the coupling between the graphene and its left and right reservoirs (we take $\Gamma^L=\Gamma^R=5$ meV), $D_c = (4\pi v_F2\rho)^{1/2}=3$ eV is an upper energy limit, $v_F$ is the Fermi velocity of graphene, and $\rho$ is its planar density \cite{PhysRevB.73.125411}. Crucially, Eq. (\ref{eq-diddv}) indicates that the localization correction to the differential conductance decreases with increasing voltage, exhibiting a logarithmic scaling as a function of this parameter that is highly suggestive of that found in our experiment.

\begin{figure}[t]
\begin{center}
\includegraphics[width=0.99\columnwidth]{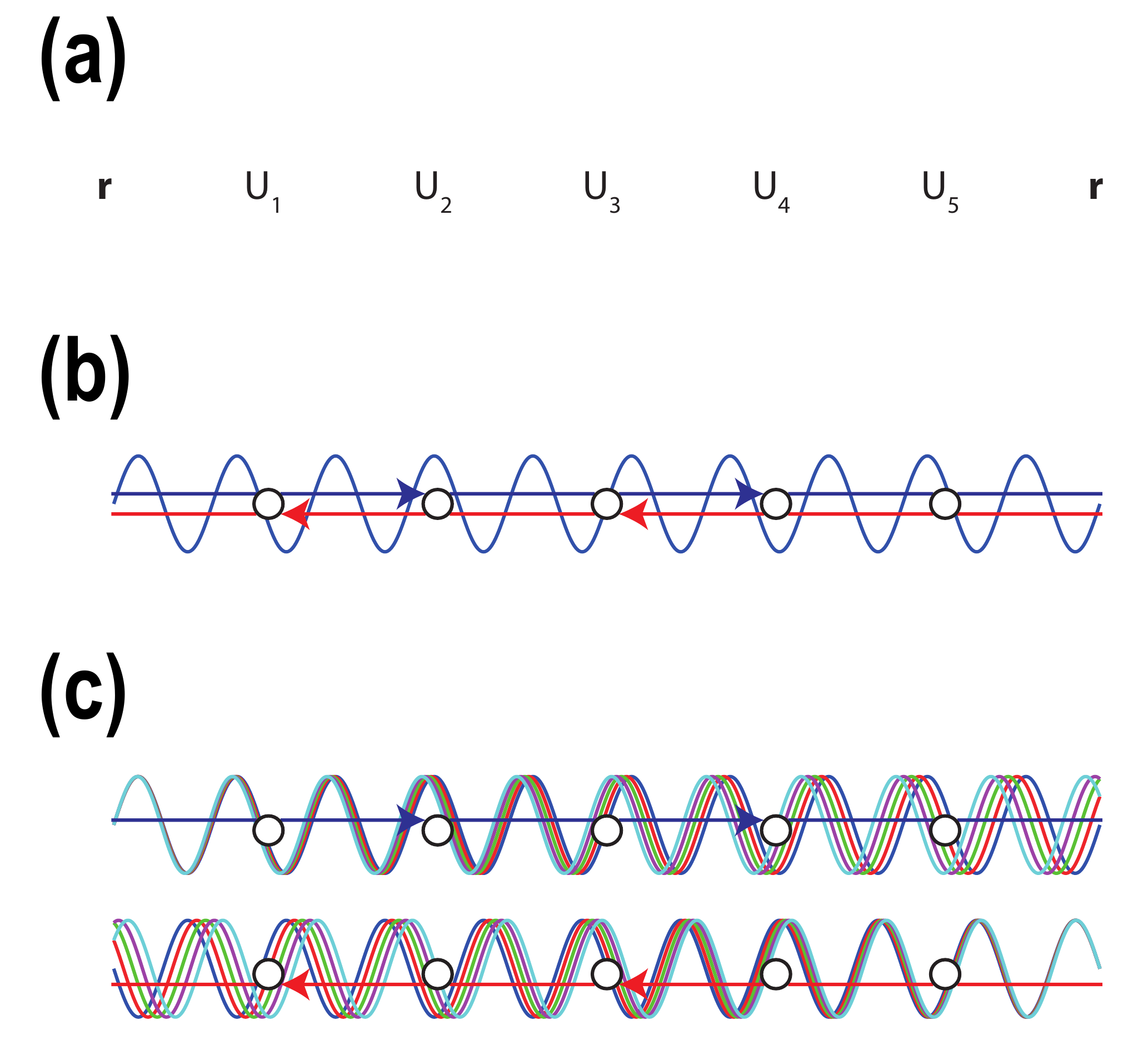}
\end{center}
\caption{A schematic illustration of the energy averaging of weak localization due to a nonzero voltage. (a) Schematic representation of a sequence of scattering events contributing to weak localization. An electron starts at point r and returns to the same point after a sequence of (five) scattering events. $U_{1-5}$ denote the different scattering potentials associated with these scatterers. (b) At zero voltage, weak localization arises from the interference of monoenergetic partial waves that travel in opposite directions. (c) At non-zero voltage, the monoenergetic waves in (b) are replaced by a set of waves with a spread of energies, which decohere with one another over an effective, voltage-induced, decoherence length.}
\label{Fig5}
\end{figure}

Before undertaking a quantitative comparison of our experiment and theory, we comment on the significance of Eq. (\ref{eq-diddv}). Physically, this describes an additional dephasing that is introduced in the transport, and which arises from a self-averaging effect whose essential idea is as follows. At zero temperature, and in the absence of any magnetic field or applied voltage, there is a fixed phase difference between electron waves that traverse any scattering loop in opposite directions. Increasing temperature leads to a summation over several such loops, each with the same fixed phase difference, so that the constructive interference that is the origin of weak localization is maintained. (This is the well-known statement that weak localization is not subject to thermal averaging \cite{RevModPhys.57.287}.) When a nonzero voltage is now applied, however, each diffusing electron essentially corresponds to a set of partial waves, with a spread of energies determined by the value of the applied voltage. As these waves diffuse through the graphene sheet, this energy spread leads to a natural dephasing, as indicated in Fig. \ref{Fig5}.

\begin{figure}[t]
\begin{center}
\includegraphics[width=0.99\columnwidth]{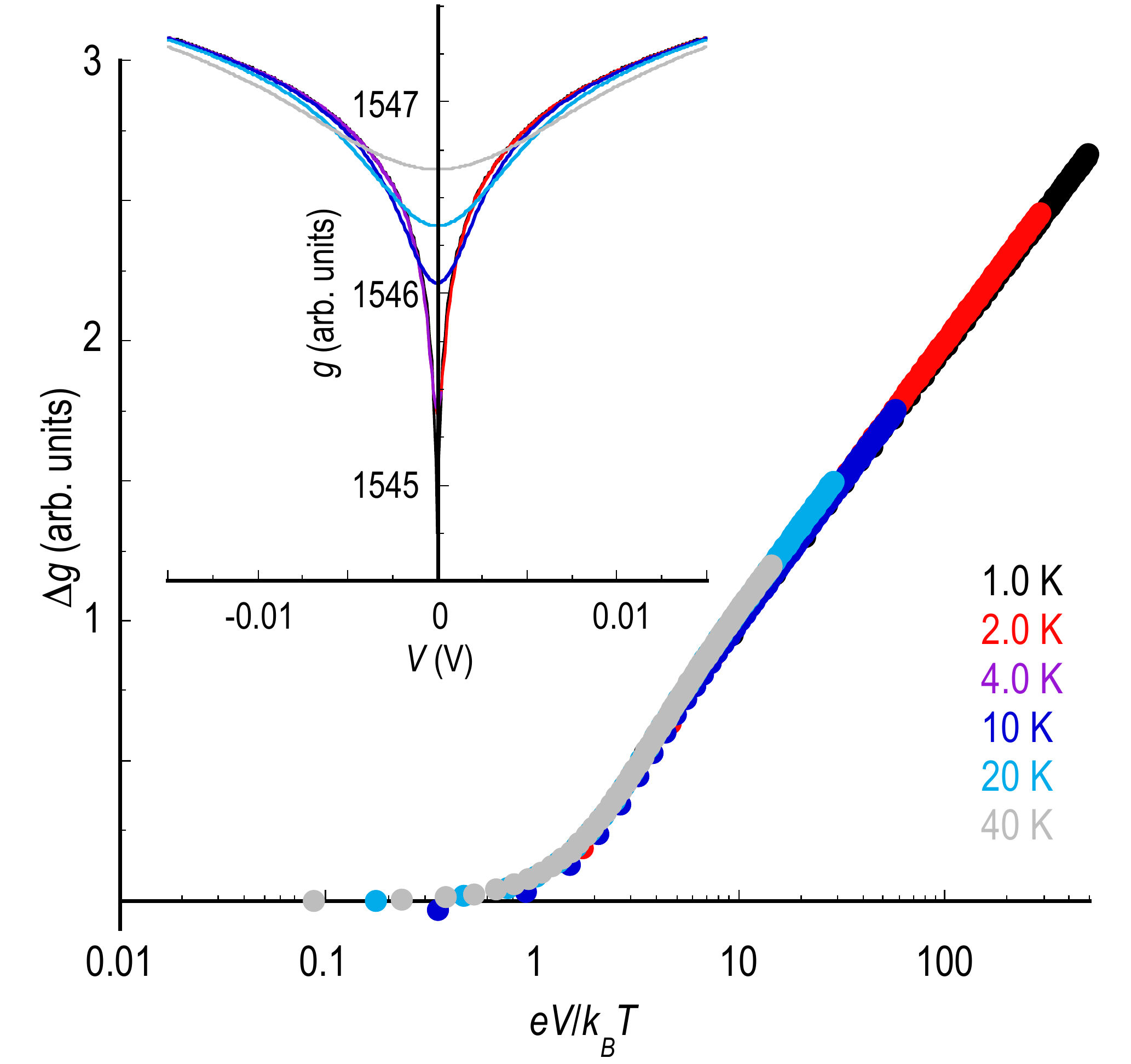}
\end{center}
\caption{In the inset to this figure, we show the calculated differential conductance for monolayer graphene (temperatures indicated) in the presence of weak localization. The impurity concentration is set at 1 \% and the scattering potential associated with the impurities is 1 eV. In the main panel we replot the differential conductance data to show the variation of bias induced conductance change ($\Delta g$) as a function of the dimensionless voltage ($eV/k_BT$).}
\label{Fig6}
\end{figure}

In the inset to Fig. \ref{Fig6}, we plot the calculated contribution of weak localization to the differential conductance of graphene, over a temperature range similar to that studied in experiment. In these calculations, the impurity concentration is taken to be 1 \% and the scattering potential associated with the impurities ($U_n = U_{1-5}$ in Fig. \ref{Fig5}) is set at 1 eV. The graphene is furthermore assumed to be intrinsic, by which we mean that the Fermi level lies at the Dirac point at thermal equilibrium. The resulting curves capture well the conductance variations found in experiment (compare, for example, with the results of Fig. \ref{Fig3}), showing a zero-bias anomaly that is suppressed with increasing voltage and temperature. The values of the characteristic voltage (10 mV) and temperature (40 K) required to suppress the anomaly are moreover consistent with the results of our experiment. To further highlight the extent of the agreement between experiment and theory, in the main panel of Fig. \ref{Fig6} we plot the result of rescaling the calculated conductance curves, using the same approach as that applied to the experimental data. Here, also, we find that the data conform to a universal voltage scaling, showing little change in differential conductance when $eV/k_BT\lesssim1$, followed by a crossover to a logarithmically increasing conductance at larger voltages. Most importantly, the collapse of the temperature-dependent data onto a single curve in this figure, without the need to make use of any adjustable parameters, reproduces the key observation of our experiment. Overall, the suggestion is that our theoretical model therefore describes the essential physics exhibited in the experiment.

\section{Discussion}
\label{sec-discussion}
The similar properties of the zero-bias anomaly, obtained in our experiment and theory, suggest a close connection of the results of Section \ref{sec-expresults} to the energy-averaging mechanism highlighted in Section \ref{sec-theory} (see Fig. \ref{Fig5} and related discussion). While we have calculated the effect of this averaging on the weak-localization correction alone, we know from our experiment (see Fig. \ref{Fig2}) that the observed conductance anomaly arises from the combined influence of localization and electron interactions. While a theoretical treatment of the latter correction lies beyond the scope of the current work, the close similarity exhibited between our experimental and theoretical results suggests, at least, that the interactions should be subject to a similar energy averaging. Indeed, in a previous study of the differential conductance of metallic nanobridges, the observed zero-bias anomaly was attributed to a quenching of the interaction correction alone, rather than the influence of weak localization \cite{PhysRevLett.70.841,PhysRevB.63.165426}. Our results for graphene clearly point to the combined influence of both mechanisms.

The role of energy-averaging in mesoscopic transport has been highlighted previously, in discussions of the thermal damping of universal conductance fluctuations in dirty metals \cite{PhysRevB.35.1039}. There, the averaging is described as an unavoidable source of static dephasing, arising from the thermal spread of the electron energy at nonzero temperature. In such systems, this static dephasing must be considered in addition to the dynamic dephasing \cite{JPCM.14.R501}, generated by scattering from time-dependent sources (such as phonons and other electrons). One question that arises here is whether the observed features of the conductance anomaly really do result from energy averaging? An alternative scenario that might be considered is that application of the bias voltage instead increases dynamic dephasing, by enhancing electron-phonon and electron-electron scattering. This would likely lead to a power-law variation of the dynamic dephasing length ($l_\varphi\propto V^{-p}$ , $p\sim1$, \cite{JPCM.14.R501}), and thus, once again, to a logarithmic conductance scaling. In such a situation, however, it is not at all clear that the resulting conductance variations should exhibit the universal voltage scaling found here. Specifically, the observation that the conductance curves collapse on one another when rescaled by the thermal energy, and that the rescaled data show a crossover in behavior when $eV\sim k_BT$, would appear to favor the energy-averaging picture presented here.

In usual discussions of weak localization \cite{RevModPhys.57.287,PhysRevB.35.1039}, the static dephasing associated with thermal smearing is known to leave the localization unaffected. Physically, this result that may be understood in the following way. At nonzero temperature, the backscattering responsible for weak localization can be attributed to a family of partial waves, with an energy spread set by the temperature. At each fixed energy within this range, a contribution to localization arises from the interference of time-reversed pairs of closed trajectories. Since each such pair returns to its origin in phase, summing over all relevant energies preserves the coherence required for the localization correction, and there is no thermal averaging of this effect. In our calculations of the energy averaging induced under nonequilibrium, however, the presence of a nonzero voltage does lead to averaging of the localization effect. Formally, this is due to the different way in which the contributions of the various scattering diagrams are summed in the nonequilibrium model. Specifically, at nonzero voltage, the localization correction must be viewed as arising from the interference of a set of partial waves that traverse different closed loops, and which have a spread of energies set by the voltage. As the voltage is increased and the size of this energy window grows, this leads to self-averaging of the wavefunction as diffusing waves gradually decohere with one another as they propagate around the same loop (see Fig. \ref{Fig5}). Mathematically, this means that the energy averaging is described by first calculating the contribution of each scattering loop for a spread of energies, and only then adding the contributions of these averaged terms.

Finally, we note that our calculations of the localization contribution are performed for intrinsic graphene, by which we mean that the Fermi level is taken to lie at the Dirac point at thermal equilibrium. As demonstrated in Fig. \ref{Fig4}, however, the bias-induced suppression of the quantum corrections, and the zero-bias anomaly that it leads to, appears to be a general feature of mesoscopic transport, for both electrons and holes in graphene. Similarly, while our model is formulated for a monolayer sheet of graphene, the observation of similar zero-bias anomalies in our monolayer and bilayer devices point to the more general nature of the voltage-induced energy averaging.


\section{Conclusions}
\label{sec-conclusions}
In conclusion, in this work we have explored the manner in which the quantum corrections to the low-temperature conductance of graphene, arising from weak localization and from electron interactions, are modified under nonequilibrium. In our studies of the differential conductance of monolayer and bilayer devices, we have demonstrated the presence of a zero-bias anomaly at low temperatures, which arises from a voltage-induced averaging of the quantum corrections. By implementing a simple rescaling of these data, in which we plot the bias-induced change of differential conductance as a function of the dimensionless voltage ($eV/k_BT$), we have shown how this anomaly collapses onto a universal, temperature-independent form. According to this, the linear conductance remains approximately unchanged for voltages $eV\lesssim k_BT$, while at larger voltages it increases as a logarithmic function of $V$, reflecting the quenching of the quantum corrections. For insight into the origins of this behavior, we have made use of nonequilibrium Green functions to formulate a formal description of weak-localization in the presence of nonlinear transport. According to this model, the voltage applied under nonequilibrium gives rise to an additional dephasing in transport, arising from a self-averaging effect. By establishing the manner in which the quantum corrections are suppressed in graphene, our study will be of broad relevance to the investigation of nonequilibrium transport in mesoscopic systems in general. This includes systems implemented from conventional metals and semiconductors, as well as those realized using other two-dimensional semiconductors \cite{NatureComms.6.7702, PhysRevB.94.245404} and topological insulators \cite{PRL.108.036805,NanoLett.13.48}.

\emph{Acknowledgements}. JF was supported by Vetenskapsrådet and collaborated on the theoretical component of this study with JPB, who is grateful for support from the National Science Foundation (ECCS-1509221). HR and GH performed device fabrication and acknowledge support from the U.S. Department of Energy, Office of Basic Energy Sciences, Division of Materials Sciences and Engineering (DE-FG02-04ER46180). RS performed differential-conductance measurements and acknowledges support from King Mongkut' s Institute of Technology Ladkrabang (contract number KREF046102).


\appendix
\section{Model and transport formalism}
\label{sec-model}
We model the low-energy physics around the Fermi level of pristine graphene using the tight-binding Hamiltonian
\begin{align}
\Hamil_0=&
	-t\sum_{\av{ij}\sigma}a^\dagger_{i\sigma}b_{j\sigma}+H.c. ,
\end{align}
where $a_{i\sigma}$ and $b_{j\sigma}$ denote the electron operators in the $A$- and $B$- sub-lattice, respectively. The nearest neighbor ($\av{ij}$) intersite hopping rate is denoted by $t$. Assuming a spin-degenerate system, we can drop the spin subscript $\sigma=\up,\down$.
We add a dilute random dispersion of impurities through
\begin{align}
\Hamil_I=&
	\int\Psi^\dagger(\bfr)\bfV(\bfr)\Psi(\bfr)d\bfr,
\end{align}
where $\Psi(\bfr)=\int\Psi_\bfk e^{-i\bfk\cdot\bfr}d\bfk/\rho$, $\Psi_\bfk=(a_\bfk\ b_\bfk)^t$, $\rho$ is the graphene planar density \cite{PhysRevB.73.125411}, and $\bfV(\bfr)=\sum_m\bfV_m\delta(\bfr-\bfr_m)$ denotes the scattering potential due to these impurities. Here, $\bfV_m=U\sum_m(\sigma_A\mathbb{1}_{m\in A}+\sigma_B\mathbb{1}_{m\in B})$ with $\sigma_A=(\sigma_0+\sigma_z)/2$ ($\sigma_B=(\sigma_0-\sigma_z)/2$), $\mathbb{1}_{m\in A(B)}$ is the indicator function for $\bfr_m$ within the $A$-sublattice ($B$-sublattice), and $U$ is the scattering potential.
%

In reciprocal space, the model $\Hamil_\text{gr}=\Hamil_0+\Hamil_I$ is transformed into $\Hamil_\text{gr}=\sum_\bfk\Psi^\dagger_\bfk\Phi_\bfk\Psi_\bfk+\sum_{\bfk\bfk'}\Psi^\dagger_\bfk\bfV_{\bfk\bfk'}\Psi_{\bfk'}$. Here
\begin{align}
\Phi_\bfk=&
	\begin{pmatrix}
		0 & \phi_\bfk \\
		\phi_\bfk^* & 0
	\end{pmatrix}
	,
\end{align}
where the structure factor $\phi_\bfk=-t\sum_j\exp(i\bfk\cdot\bfdelta_j)$ is given in terms of the nearest neighbor vectors $\bfdelta_1=a(\sqrt{3},1)/2$, $\bfdelta_2=-a(\sqrt{3},-1)/2$, and $\bfdelta_3=-a(0,1)$, with lattice parameter $a$. Free electrons have the dispersion relation $\phi_{\bfk\pm\bfK}\approx\pm v_Fke^{\pm i\varphi}$ around the $K$-points $\bfK_\pm=\pm\bfK=\pm4\pi\sqrt{3}(1,0)/9a$, with Fermi velocity $v_F=3at/2$.
In $\bfk$-space, the scattering potential $\bfV_{\bfk\bfk'}=\sum_m\bfV_m\exp[-i(\bfk-\bfk')\cdot\bfr_m]/\Omega$, where $\Omega$ is the volume.

The conductance of the disordered graphene flake is calculated by placing it in the junction between a pair of metallic leads, modeled here with Hamiltonians $\Hamil_L=\sum_\bfp(\dote{\bfp}-\mu_L)\csdagger{\bfp}\cs{\bfp}$ and $\Hamil_R=\sum_\bfq(\dote{\bfp}-\mu_R)\csdagger{\bfq}\cs{\bfq}$, where the chemical potentials $\mu_{L/R}$ are related to the applied voltage $V$ by $\mu_L-\mu_R=eV$. Tunneling between the leads and the graphene is described by the Hamiltonian $\Hamil_T=\sum_{\bfp\bfk}\csdagger{\bfp}t_{\bfp\bfk}\Psi_{\bfk}+\sum_{\bfq\bfk}\csdagger{\bfq}t_{\bfq\bfk}\Psi_{\bfk}+H.c.$, where the row vector $t_{\bfp\bfk}$ ($t_{\bfq\bfk}$) denotes the tunneling rate between the left (right) lead and the graphene. It must be kept in mind that electrons in both sub-lattices take part in tunneling to and from the leads, which is accounted for here by the vectors $t_{\bfp\bfk}$ and $t_{\bfq\bfk}$.

The stationary charge current is given by $I=-e\dt\av{N_L}=-e\dt\sum_{\bfk\sigma}\av{\cdagger{\bfp}\cc{\bfp}}$, which using standard methods becomes
\begin{align}
I=&
	\frac{ie}{h}
		\tr\sum_{\bfk\bfk'}
		\int
			\bfGamma^L_{\bfk\bfk'}
			\biggl(
				f_L(\omega)\bfG^>_{\bfk'\bfk}(\omega)
				+
				f_L(-\omega)\bfG^<_{\bfk'\bfk}(\omega)
			\biggr)
		d\omega
		.
\label{eq-I}
\end{align}
Here, the trace runs over the pseudo-spin degrees of freedom, $f_\chi(x)=f(\omega-\mu_\chi)$ is the Fermi function at the chemical potential $\mu_\chi$, and $\bfGamma^\chi_{\bfk\bfk'}$ is the coupling between the lead $\chi = L, R$ and the central region. We omit the momentum dependence of the coupling, $\bfGamma^\chi_{\bfk\bfk'}=\bfGamma^\chi$ and assume that the lesser/greater Green function for the central region $\bfG^{</>}_{\bfk\bfk'}(\omega)=\bfG^{</>}_\bfk(\omega)$.

\section{Electronic structure calculation}
\label{sec-elstruct}
We describe the weak-localization correction in graphene by considering the features of its electronic structure. In these calculations, the assumption of nonequilibrium conditions requires that we expand all calculated quantities on the Keldysh contour.

The electronic structure of pristine graphene is described by the \emph{free} (unperturbed) graphene Green function
$\bfg(\bfk;z)=(z\sigma_0+\Phi_\bfk)(z^2-|\phi_\bfk|^2)$, where $z\in\mathbb{C}$.
We write the equation of motion for the Green function $\bfG(\bfk,\bfk';z)=\av{\inner{\Psi_\bfk}{\Psi^\dagger_{\bfk'}}}(z)$ as the Dyson equation
\begin{align}
\bfG_{\bfk\bfk'}=&
	\delta_{\bfk\bfk'}\bfg_\bfk
	+
	\sum_{\bfkappa}\bfg_\bfk\bfV_{\bfk\bfkappa}\bfG_{\bfkappa\bfk'}
	,
\end{align}
and expand in orders of the scattering potential $\bfV_{\bfk\bfk'}$,
\begin{align}
\bfG_{\bfk\bfk'}=&
	\delta_{\bfk\bfk'}\bfg_\bfk
	+
	\bfg_\bfk\bfV_{\bfk\bfk'}\bfg_{\bfk'}
	+
	\sum_{\bfkappa}\bfg_\bfk\bfV_{\bfk\bfkappa}\bfg_{\bfkappa}\bfV_{\bfkappa\bfk'}\bfg_{\bfk'}
	+
	\cdots
	,
\label{eq-Vexpansion}
\end{align}
which enables an order-by-order investigation of the electronic structure in terms of the scattering potential.

\subsection{Impurity averaging}
\label{ssec-graveraging}
\begin{figure}[t]
\begin{center}
\includegraphics[width=0.99\columnwidth]{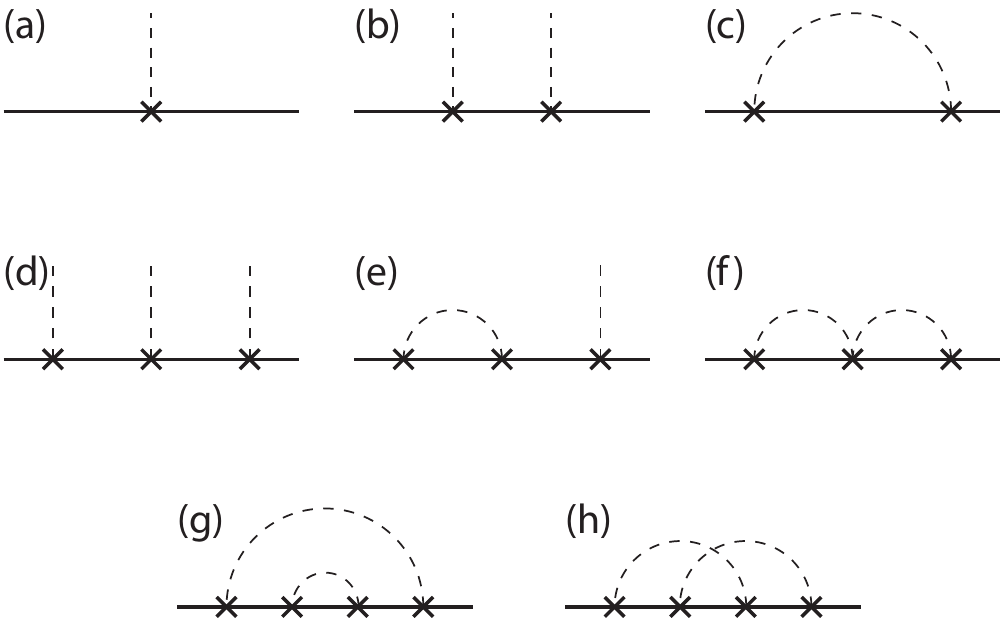}
\end{center}
\caption{Some of the low-order diagrams that are important for conductivity calculations. Momentum is conserved at each vertex. Solid (dashed) lines represent free-electron Green function (impurity potential), whereas crosses mark scattering events. Diagrams (c), (f), and (g) are the lowest-order rainbow diagrams, whereas diagram (h) is the lowest-order crossed diagram.}
\label{fig-Fig1}
\end{figure}

We calculate the weak localization correction by making an average over impurities located at $\{\bfr_m\}$, thereby surrendering the non-locality of the Green function in reciprocal space. Employing the method outlined in \cite{SolStComm.86.195,QuantumKinetics}, we obtain to first order
\begin{align}
\overline{\bfV}_{\bfk\bfk'}=&
	\frac{U}{\Omega}
		(N_A\sigma_A+N_B\sigma_B)
		\delta_{\bfk\bfk'}
	.
\end{align}
The Feynman diagram corresponding to this scattering process is depicted in Fig. \ref{fig-Fig1} (a).
Assuming equal numbers of impurities in the two sublattices, $N_A=N_B=N$, we can write
\begin{align}
\overline{\bfV}_{\bfk\bfk'}=&
	\frac{N}{\Omega}U\sigma_0\delta_{\bfk\bfk'}
	=
	cU\sigma_0\delta_{\bfk\bfk'}
	,
\end{align}
where $c=N/\Omega$ defines the concentration of impurities. The first order correction to the Green function is, hence, given by
\begin{align}
\delta\bfG_\bfk^{(1)}=&
	cU\bfg^2_\bfk.
\end{align}
In an analogous manner, the second-order component becomes
\begin{align}
\overline{\bfV_{\bfk\bfkappa}\bfg_{\bfkappa}\bfV_{\bfkappa\bfk'}}=&
	\frac{NU^2}{\Omega^2}
	\biggl(
		(N-1)\bfg_{\bfkappa}\sigma_0
		+
		\sum_{i=A,B}\sigma_i\bfg_{\bfkappa}\sigma_i
	\biggr)
	\delta_{\bfk\bfk'}.
\end{align}
The Feynman diagrams corresponding to this second-order scattering process are depicted in Figs. \ref{fig-Fig1} (b) and (c), respectively.
The distribution of impurities between the two sub-lattices leads to a restricted contribution from $\sum_{\bfkappa}\bfg_{\bfkappa}$, picking out the diagonal components only. This structural organization is important for some of the higher-order contributions.

The second-order correction is summarized as
\begin{align}
\delta\bfG_\bfk^{(2)}=&
	c^2U^2\biggl[1-\frac{1}{N}\biggr]\bfg^3_\bfk
	+
	cU^2\bfg_\bfk\tilde\bfg\bfg_\bfk
	,
\end{align}
where $\tilde\bfg=\sum_{i=A,B}\sum_{\bfkappa}\sigma_i\bfg(\bfkappa)\sigma_i$. Here, the first term can be absorbed into the single-particle energy. The last term, however, provides the first diagram in the class of rainbow diagrams, see Fig. \ref{fig-Fig1} (c). A partial summation over this class of diagrams leads to an electronic structure that is dependent upon the effective impurity-limited scattering lifetime  $\tau_I$. Therefore, we define the self-energy in the self-consistent Born approximation in terms of the impurity averaged Green function $\overline\bfG$ ($u^2=cU^2$)
\begin{align}
\bfSigma=&
	\frac{u^2}{\Omega}
	\sum_{i=A,B}
	\sum_\bfk
		\sigma_i\overline\bfG_{\bfkappa}\sigma_i,
\end{align}
where the impurity averaged Green function is given in terms of the Dyson equation $\overline\bfG_\bfk=(\bfg^{-1}(\bfk)-\bfSigma[\overline\bfG])^{-1}$. The ansatz $\bfSigma^r(\omega)=(\Lambda-i/2\tau_I)\sigma_0$ for the retarded self-energy, leads to
\begin{align}
\frac{u^2}{\Omega}
	\sum_{i=A,B}&
	\sum_\bfk
		\sigma_i\overline{\bfG}_\bfk^r\sigma_i
\nonumber\\=&
	-\frac{4u^2}{D_c^2}
	\biggl(
		\omega+\frac{i}{2\tau_I}
	\biggr)
	\biggl(
		\ln\frac{D_c}{|\omega+i/2\tau_I|}
		+
		i\frac{\pi}{2}\sign(\omega)
	\biggr)
	\sigma_0,
\end{align}
where $D_c^2=4\pi v_F^2\rho$ defines an upper energy cut.
Equating for $\Lambda$ and $1/\tau_I$,
and assuming that $1/\tau_I\ll|\omega|$ (which is sufficient for small energies around the Fermi level, since $D_c\sim1$ eV, and $U\sim1eV$, while $u^2/D_c^2\sim10^{-3}$ | $10^{-2}$, implying that $1-4(u^2/D_c^2)\ln(D_c/|\omega+i/2\tau_I|)\geq1$), leads to the result that we can neglect $\Lambda$, and retain only the approximate inverse lifetime, or, momentum scattering time
\begin{align}
\frac{1}{\tau_I}\approx&
	\frac{4\pi u^2}{D_c^2}|\omega|
	,
\label{eq-tau}
\end{align}
in agreement with previous studies \cite{JPhysSocJpn.67.2421,PhysRevLett.89.266603}.

\subsection{Maximally-crossed diagrams}

The weak localization phenomenon arises from enhanced backscattering, caused by the constructive interference between pairs of time-reversed closed paths. It is well known that diagrams containing crossed impurity lines account for this interference and that the largest contribution from this class is provided by the subclass of maximally-crossed diagrams. The diagram in Fig. \ref{fig-Fig1} (h), which represents the lowest-order maximally-crossed diagram and can be written algebraically as 
\begin{align}
\bfSigma^\text{(1)}_{cr}(\bfk)=&
	\frac{1}{\Omega^4}
	\sum_{m\neq n}
	\sum_{\bfq\bfkappa}
		\bfV_m
		\overline\bfG_{\bfp/2+\bfkappa}
		\bfV_n
		\overline\bfG_\bfq
		\bfV_m
		\overline\bfG_{\bfp/2-\bfkappa}
		\bfV_n
\nonumber\\\approx&
	\frac{u^4}{\Omega^2}
	\sum_{\bfq\bfkappa}
	\sum_{i,j=A,B}
		\sigma_i
		\overline\bfG_{\bfp/2+\bfkappa}
		\sigma_j
		\overline\bfG_\bfq
		\sigma_i
		\overline\bfG_{\bfp/2-\bfkappa}
		\sigma_j
	,
\label{eq-Scr}
\end{align}
where $\bfp=\bfk+\bfq$.
%
The dominant contribution to the lesser/greater form of the self-energy is then given by \cite{QuantumKinetics}
\begin{align}
\bfSigma^{\text{(1)}</>}_{cr}(\bfk)=&
	\frac{u^4}{\Omega^2}
	\sum_{\bfq\bfkappa}
	\sum_{i,j=A,B}
		\sigma_i
		\overline\bfG_{\bfp/2+\bfkappa}^r
		\sigma_j
		\overline\bfG_\bfq^{</>}
		\sigma_i
		\overline\bfG_{\bfp/2-\bfkappa}^a
		\sigma_j
	.
\end{align}


The matrices $\sigma_i$, $i=A,B$, are projections and orthogonal ($\sigma_i\sigma_j=\delta_{ij}\sigma_i$). For a general $2\times2$ matrix $\bfA$, we have $\sigma_A\bfA\sigma_A=A_{11}\sigma_A$ and $\sigma_B\bfA\sigma_B=A_{22}\sigma_B$, while $\sigma_A\bfA\sigma_B=A_{12}\sigma_+$ and $\sigma_B\bfA\sigma_A=A_{21}\sigma_-$, where $\sigma_\pm=(\sigma_x\pm i\sigma_y)/2$. Hence, a product on the form $\sigma_i\bfA\sigma_j\bfB\sigma_i\bfC\sigma_j$ reduces to
\begin{align}
\sum_{i,j=A,B}\sigma_i\bfA\sigma_j\bfB\sigma_i\bfC\sigma_j=&
	\begin{pmatrix}
		A_{11}B_{11}C_{11} & A_{12}B_{21}C_{12} \\
		A_{21}B_{12}C_{21} & A_{22}B_{22}C_{22} 
	\end{pmatrix}
	.
\label{eq-matrixstructure}
\end{align}
This property enables us to evaluate all of the crossed diagrams, element by element.

Since the diagonal components of the graphene Green function are equal, we only have to consider one of them. An analogous observation holds for the off-diagonal components. The calculation of all matrix components in the self-energy is fundamentally important since the coupling between the pseudo-spin degrees of freedom plays a central role in the theory of weak localization in graphene \cite{PhysRevLett.97.146805}. Hence, we write
\begin{align}
\Bigl(
	\bfSigma^{(1)</>}_{cr}(\bfk)\Bigr)_{ij}=&
	\frac{u^2}{\Omega}\sum_\bfq
		\zeta_{ij}(\bfp)
		\Bigl(\overline{\bfG}^{</>}_\bfq\Bigr)_{ji}
	,
\end{align}
where the subscripts $ij$ refer to matrix components and where
\begin{align}
\zeta_{ij}(\bfp)=&
	\frac{u^2}{\Omega}
	\sum_{\bfkappa}
		\Bigl(\overline{\bfG}^r_{\bfp/2+\bfkappa}\Bigr)_{ij}
		\Bigl(\overline{\bfG}^a_{\bfp/2-\bfkappa}\Bigr)_{ij}
	.
\end{align}

The momentum summation contained in $\zeta_{ij}$ is carried out by individually expanding the Green functions around the nodes in the two valleys $\pm\bfK$, such that both intra- and inter-valley scattering is included. Hence, 
\begin{align}
\phi_{\bfp/2\pm\bfkappa+s\bfK}\approx&
	s\frac{v_F}{2}(pe^{is\varphi_\bfp}\pm2\kappa e^{is\varphi_{\bfkappa}})
	,\ s=\pm,
\end{align}
where $\tan\varphi_\bfk=k_y/k_x$ and  $\tan\varphi_\bfp=p_y/p_x$. This leads to the result that
\begin{align}
\phi_{\bfp/2+\bfkappa}\phi_{\bfp/2-\bfkappa}\approx&
	-E^2\sin^2\varphi_\bfp+4\dote{}^2\sin^2\varphi_{\bfkappa}
	,
\end{align}
where $E=v_Fp$ and $\dote{}=v_F\kappa$. Moreover, since $|\phi_{\bfp/2\pm\bfkappa}|\rightarrow v_F|\bfp/2\pm\bfkappa|$ in the valleys, and restricting ourselves to the regime $E\ll\dote{F}$, we can employ the approximation $v_F^2|\bfp/2\pm\bfkappa|^2\approx\dote{}^2\pm\dote{F}E\cos\gamma$. In the notation $(z^{r/a}_\mp)^2=(z^{r/a})^2\mp\dote{F}E\cos\gamma$, with $z^{r/a}=\omega\pm i/2\tau$ and $\gamma=\varphi_{\bfkappa}-\varphi_\bfp$, we have
\begin{subequations}
\begin{align}
\zeta_{11}(\bfp)=&
	\frac{8u^2}{D_c^2}
	\int
		\frac{|z^r|^2}
		{[(z^r_-)^2-\dote{}^2][(z^a_+)^2-\dote{}^2]}
	\frac{\dote{}d\dote{}d\varphi_{\bfkappa}}{2\pi}
	,
\\
\zeta_{12}(\bfp)=&
	-\frac{2u^2}{D_c^2}
	\int
		\frac{E^2\sin^2\varphi_\bfp-4\dote{}^2\sin^2\varphi_{\bfkappa}}
		{[(z^r_-)^2-\dote{}^2][(z^a_+)^2-\dote{}^2]}
	\frac{\dote{}d\dote{}d\varphi_{\bfkappa}}{2\pi}
	.
\label{eq-zeta12}
\end{align}
\end{subequations}
The energy ($\dote{}$) integration generates the contribution $\log[D_c^2/(iz^r_-)^2]-\log[D_c^2/(iz^a_+)^2]\approx i2\pi$ in both integrals, while $\zeta_{12}$ also contains the contribution
\begin{align}
\log
	\frac{D_c^4}{(z^r_-)^2(z^a_+)^2}
	\approx&
	4\ln\frac{D_c}{|\omega|}
	.
\label{eq-lnDcw}
\end{align}
In these equations, we have omitted any angular dependence in the logarithms, which in the latter case implies that $(z^r_-)^2(z^a_+)^2\propto\omega^4$.

We can now write the kernels $\zeta_{ij}$ according to
\begin{subequations}
\begin{align}
\zeta_{11}(\bfp)=&
	i\frac{2\pi u^2}{D_c^2}
	\int
		\frac{|z^r|^2}
			{i\omega/\tau-\dote{F}E\cos\gamma}
	\frac{d\varphi_{\bfkappa}}{2\pi}
	,
\\
\zeta_{12}(\bfp)=&
	-\frac{u^2}{D_c^2}
	\int
	\biggl(
		4\sin^2\varphi_{\bfkappa}\ln\frac{D_c}{|\omega|}
\nonumber\\&
		-
		i\pi
		\frac{E^2\sin^2\varphi_\bfp-4\omega^2\sin^2\varphi_{\bfkappa}}
			{i\omega/\tau-\dote{F}E\cos\gamma}
	\biggr)
	\frac{d\varphi_{\bfkappa}}{2\pi}
	.
\end{align}
\end{subequations}
Then, by performing the angular integration, we arrive at
\begin{subequations}
\begin{align}
\zeta_{11}(\bfp)=&
	\frac{2\pi u^2}{D_c^2}
		\frac{\omega^2}
			{\sqrt{(\omega/\tau)^2+(\dote{F}E)^2}}
	,
\\
\zeta_{12}(\bfp)=&
	-\frac{u^2}{D_c^2}
	\biggl(
		2\ln\frac{D_c}{|\omega|}
		-
		\pi
		\frac{E^2-4\omega^2}
			{\sqrt{(\omega/\tau)^2+(\dote{F}E)^2}}
		\sin^2\varphi_\bfp
	\biggr)
	,
\end{align}
\end{subequations}
where in the result for $\zeta_{11}$ we have replaced $|z^r|$ by $\omega$ (since $1/4\tau^2\ll\omega^2$) and in that for $\zeta_{12}$ we have neglected a minor contribution proportional to $\omega^2$. In fact, the second contribution to $\zeta_{12}$ can also be omitted without changing its vital properties pertaining to weak localization.

To first order in the self-energy, our calculation of the maximally-crossed diagrams yields the expression
\begin{align}
\bfSigma^{(1)</>}_{cr}(\bfk)=&
	\frac{u^2}{\Omega}
		\sum_\bfq
		\left\{
			\frac{2\pi u^2\omega^2/D_c^2}{\sqrt{(\omega/\tau)^2+(\dote{F}E)^2}}
			\begin{pmatrix}
				\Bigl(\overline{\bfG}^{</>}_\bfq\Bigr)_{11} & 0 \\
				0 & \Bigl(\overline{\bfG}^{</>}_\bfq\Bigr)_{22}
			\end{pmatrix}
		\right.
\nonumber\\&
		\left.
			-
			\frac{2u^2}{D_c^2}\ln\frac{D_c}{|\omega|}
			\begin{pmatrix}
				0 & \Bigl(\overline{\bfG}^{</>}_\bfq\Bigr)_{21} \\
				\Bigl(\overline{\bfG}^{</>}_\bfq\Bigr)_{12} & 0
			\end{pmatrix}
		\right\}
		.
\end{align}
The result in Eq. (\ref{eq-matrixstructure}) implies that the $n^{th}$ contribution to both diagonal and off-diagonal entries equal the $n^{th}$ power of the entries in the expression above. Hence, the summation over the maximally-crossed diagrams can be performed independently, in each matrix entry, giving
\begin{subequations}
\begin{align}
\sum_{n=1}^\infty
	\Bigl(\zeta_{11}(\bfp)\Bigr)^n
	=
	\sum_{n=1}^\infty
	\Bigl(\zeta_{22}(\bfp)\Bigr)^n
	\approx&
	\frac{(\omega/\tau\dote{F})^2}{E^2+(\omega/\tau\dote{F})^2}
	,
\\
\sum_{n=1}^\infty
	\Bigl(\zeta_{12}(\bfp)\Bigr)^n
	=
	\sum_{n=1}^\infty
	\Bigl(\zeta_{21}(\bfp)\Bigr)^n
	\approx&
	-
		\frac{2u^2}{D_c^2}
		\ln\frac{D_c}{|\omega|}
	.
\end{align}
\end{subequations}
In the expressions for $\zeta_{11}$  and $\zeta_{22}$, we have made use of the relation $4\pi u^2/D_c^2\approx1/\tau|\omega|$. The self-energy contribution from the maximally-crossed diagrams is then reduced to the expression
\begin{align}
\bfSigma^{</>}_{cr}(\bfk)=&
	\frac{u^2}{\Omega}
		\sum_\bfq
		\left\{
			\frac{(\omega/\tau\dote{F})^2}{E^2+(\omega/\tau\dote{F})^2}
			\begin{pmatrix}
				\Bigl(\overline{\bfG}^{</>}_\bfq\Bigr)_{11} & 0 \\
				0 & \Bigl(\overline{\bfG}^{</>}_\bfq\Bigr)_{22}
			\end{pmatrix}
		\right.
\nonumber\\&
		\left.
			-
			\frac{2u^2}{D_c^2}
			\ln\frac{D_c}{|\omega|}
			\begin{pmatrix}
				0 & \Bigl(\overline{\bfG}^{</>}_\bfq\Bigr)_{21} \\
				\Bigl(\overline{\bfG}^{</>}_\bfq\Bigr)_{12} & 0
			\end{pmatrix}
		\right\}
		.
\end{align}

Next, we sum over the momentum $\bfq$. In the diagonal components, we note that $1/[E^2+(\tau\dote{F}/\omega)^2]$ is strongly peaked around $\bfq=-\bfk$, and recall that $E=v_F|\bfk+\bfq|$, which allows us to move $(\overline{\bfG}^{</>}_\bfq)_{11(22)}$ out of the summation, the remainder of which yields a factor $\approx u^2$.
The summation of the off-diagonal components is trivial since the contribution from $\zeta_{12}$ is independent of $\bfq$. Setting $\overline{\bfG}^{</>}_\bfq=\overline{\bfG}^r_\bfq\bfSigma^{</>}_0\overline{\bfG}^a_\bfq$, where $\bfSigma^{</>}_0(\omega)=(\pm i)\sum_\chi\Gamma^\chi f_\chi(\pm\omega)(\sigma_0+\sigma_x)$ leads to
\begin{align}
\bfSigma^{</>}_{cr}(\bfk)\approx&
			(\pm i)\sum_\chi\Gamma^\chi f_\chi(\pm\omega)
			\Biggl[
				\sigma_0
				u^2
				\biggl|
					\Bigl(\overline{\bfG}_{-\bfk}^r\Bigr)_{11}
					+
					\Bigl(\overline{\bfG}_{-\bfk}^r\Bigr)_{12}
				\biggr|^2
\nonumber\\&
				-
				\sigma_x
				\frac{3u^2}{D_c^2}
				\ln\frac{D_c}{|\omega|}
			\Biggr]
		.
\end{align}

The diagonal components in this expression provide a contribution to the current that is quartic in the Green functions, which should be compared with the quadratic contribution generated by the off-diagonal terms. Since the multiplying factor is also constant, when compared to the logarithmic functions in the off-diagonal components, the diagonal components can be discarded when calculating the weak-localization correction.
The off diagonal components account for the coupling between electrons in different sublattices, indicating that the pseudo-spin chirality is of great importance for the emergence of weak localization in graphene.

Finally, calculation of the differential conductance reduces to the procedure of taking the derivative of the relevant Fermi functions, yielding the correction
\begin{align}
\frac{d\delta I}{dV}=&
	-\frac{2e^2}{h}\Gamma^L\Gamma^R
	\int
		\calF_V(\omega)
		\frac{1}{D_c^2}
		\ln\frac{D_c}{|\omega|}
	d\omega
	.
\label{eq-weak localizationcorrection}
\end{align}
At low temperatures, this displays the logarithmic variation normally associated with weak localization in linear transport, since $\calF_V(\omega)$ $\{=(\beta/4)\cosh^{-2}\beta(\omega-eV)/2\}$ $\rightarrow\delta(\omega-eV)$, as $T\rightarrow0$, giving
\begin{align}
\frac{d\delta I}{dV}\rightarrow&
	-\frac{2e^2}{h}
	\cdot
	\frac{\Gamma^L\Gamma^R}{D_c^2}
	\ln\frac{D_c}{|eV|}
	,\ T\rightarrow0.
\end{align}


\begin{thebibliography}{99}

\bibitem{PhysRevB.28.2914} G. Bergmann, Phys. Rev. B {\bf 28}, 2914 (1983).
\bibitem{PhysRep.107.1} G. Bergmann, Phys. Rep. {\bf 107}, 1 (1984).
\bibitem{RevModPhys.57.287} P. A. Lee and T. V. Ramakrishnan, Rev. Mod. Phys. {\bf 57}, 287 (1985).
\bibitem{PhysRevLett.44.1288} B. L. Altshuler, A. G. Aronov, and P. A. Lee, Phys. Rev. Lett. {\bf 44}, 1288 (1980).
\bibitem{PhysRevB.35.4205} G. Bergmann, Phys. Rev. B {\bf 35}, 4205 (1987).

\bibitem{PhysRevLett.97.146805} E. McCann, K. Kechedzhi, V. I. Fal\'{}ko, H. Suzuura, T. Ando, and B. L. Altshuler, Phys. Rev. Lett. {\bf 97}, 146805 (2006).
\bibitem{PhysRevLett.97.036802} D. V. Khveshchenko, Phys. Rev. Lett. {\bf 97}, 036802 (2006).
\bibitem{PhysRevLett.97.196804} A. F. Morpurgo and F. Guinea, Phys. Rev. Lett. {\bf 97}, 196804 (2006).
\bibitem{PhysRevLett.98.176806} K. Kechedzhi, V. I. Fal\'{}ko, E. McCann, and B. L. Altshuler, Phys. Rev. Lett. {\bf 98}, 176806 (2007).
\bibitem{PhysRevLett.101.126801} X.-Zh. Yan and C. S. Ting, Phys. Rev. Lett. {\bf 101}, 126801 (2008).

\bibitem{SolStComm.151.1550} M.O. Nestoklon, N.S. Averkiev, S.A. Tarasenko, Sol. State Comm. {\bf 151}, 1550 (2011).
\bibitem{PhysRevLett.108.166606} E. McCann and V. I. Fal\'{}ko, Phys. Rev. Lett. {\bf 108}, 166606 (2012).
\bibitem{PhysRevLett.97.016801} S. V. Morozov, K. S. Novoselov, M. I. Katsnelson, F. Schedin, L. A. Ponomarenko, D. Jiang, and A. K. Geim, Phys. Rev. Lett. {\bf 97}, 016801 (2006).
\bibitem{PhysRevLett.98.136801} X. Wu, X. Li, Z. Song, C. Berger, and W. A. de Heer, Phys. Rev. Lett. {\bf 98}, 136801 (2007).
\bibitem{PhysRevLett.98.176805} R. V. Gorbachev, F. V. Tikhonenko, A. S. Mayorov, D. W. Horsell, and A. K. Savchenko, Phys. Rev. Lett. {\bf 98}, 176805 (2007).

\bibitem{PhysRevLett.100.056802} F. V. Tikhonenko, D. W. Horsell, R. V. Gorbachev, and A. K. Savchenko, Phys. Rev. Lett. {\bf 100}, 056802 (2008).
\bibitem{PhysRevB.78.125409} D.-K. Ki, D. Jeong, J.-H. Choi, H.-J. Lee, and K.-S. Park, Phys. Rev. B {\bf 78}, 125409 (2008).
\bibitem{ApplPhysLett.93.122102} T. Shen, Y. Q. Wu, M. A. Capano, L. P. Rokhinson, L. W. Engel, and P. D. Ye, Appl. Phys. Lett. {\bf 93}, 122102 (2008).
\bibitem{PhysRevLett.103.226801} F. V. Tikhonenko, A. A. Kozikov, A. K. Savchenko, and R. V. Gorbachev, Phys. Rev. Lett. {\bf 103}, 226801 (2009).
\bibitem{NewJPhys.11.095021} J. Eroms and D. Weiss, New J. Phys. {\bf 11}, 095021 (2009).

\bibitem{Nanotechnol.21.274014} J. Berezovsky and R. M. Westervelt, Nanotechnol. {\bf 21}, 274014 (1998).
\bibitem{JPCM.22.205301} Y.-F. Chen, M.-H. Bae, C. Chialvo, T. Dirks, A. Bezryadin and N. Mason, J. Phys.: Condens. Matt. {\bf 22}, 205301 (2010).
\bibitem{NatMater.10.443} Q. Yu, L. A. Jauregui, W. Wu, R. Colby, J. Tian, Z. Su, H. Cao, Z. Liu, D. Pandey, D. Wei, T. F. Chung, P. Peng, N. P. Guisinger, E. A. Stach, J. Bao, S.-S. Pei, and Y. P. Chen, Nat. Mater. {\bf 10}, 443 (2011).
\bibitem{ApplPhysLett.103.143111} F. Oberhuber, S. Blien, S. Heydrich, F. Yaghobian, T. Korn, C. Schuller, C. Strunk, D. Weiss, and J. Eroms, Appl. Phys. Lett. {\bf 103}, 143111 (2013).
\bibitem{JPhysSocJpn.67.2857} T. Ando, T. Nakanishi, and R. Saito, J. Phys. Soc. Jpn. {\bf 67}, 2857 (1998).

\bibitem{JPhysSocJpn.74.777} T. Ando, J. Phys. Soc. Jpn. {\bf 74}, 777 (2005).
\bibitem{PhysRevB.82.075424} A. A. Kozikov, A. K. Savchenko, B. N. Narozhny, and A. V. Shytov, Phys. Rev. B {\bf 82}, 075424 (2010).
\bibitem{PhysRevB.83.195417} B. Jouault, B. Jabakhanji, N. Camara, W. Desrat, C. Consejo, and J. Camassel, Phys. Rev. B {\bf 83}, 195417 (2011).
\bibitem{NewJPhys.13.113005} W. Pan, A. J. Ross III, S. W. Howell, T. Ohta, T. A. Friedmann and C.-T. Liang, New J. Phys. {\bf 13}, 113005 (2011).
\bibitem{PhysRevLett.108.106601} J. Jobst, D. Waldmann, I. V. Gornyi, A. D. Mirlin, amd H. B. Weber, Phys. Rev. Lett. {\bf 108}, 106601 (2012).

\bibitem{PhysRevB.88.235406} A. Iagallo, S. Tanabe, S. Roddaro, M. Takamura, H. Hibino, and S. Heun, Phys. Rev. B {\bf 88}, 235406 (2013).
\bibitem{PhysRevB.90.035423} B. Jabakhanji, D. Kazazis, W. Desrat, A. Michon, M. Portail, and B. Jouault, Phys. Rev. B {\bf 90}, 035423 (2014).
\bibitem{PhysRevLett.43.721} G.J. Dolan and D.D. Osherof, Phys. Rev. Lett. {\bf 43}, 721 (1979).
\bibitem{PhysRevB.25.5563} H. Hoffman, F. Hofmann and W. Schoepe, Phys. Rev. B {\bf 25}, 5563 (1982).
\bibitem{PhysRevB.55.4061} H. Linke, P. Omling, H. Xu, and P. E. Lindelof, Phys. Rev. B {\bf 55}, 4061 (1997).

\bibitem{PhysRevLett.70.841} H. B. Weber, R. Haussler, and E. Langheinrich, Phys. Rev. Lett. {\bf 70}, 841 (1993).
\bibitem{PhysRevB.63.165426} U. Murek, R. Schafer, and H. v. Lohneysen, Phys. Rev. B {\bf 63}, 165426 (2001).
\bibitem{PhysRevB.82.045411} S. Lee, N. Wijesinghe, C. Diaz-Pinto, and H. Peng, Phys. Rev. B {\bf 82}, 045411 (2010).
\bibitem{PhysRevB.83.205421} J. K. Viljas, A. Fay, M. Wiesner, and P. J. Hakonen, Phys. Rev. B {\bf 83}, 205421 (2011).
\bibitem{PhysRevB.84.245427} A. Fay, R. Danneau, J. K. Viljas, F. Wu, M. Y. Tomi, J. Wengler, M. Wiesner, and P. J. Hakonen, Phys. Rev. B {\bf 84}, 245427 (2011).

\bibitem{PhysRevB.85.161411} A. S. Price, S. M. Hornett, A. V. Shytov, E. Hendry, and D. W. Horsell, Phys. Rev. B {\bf 85}, 161411(R) (2012).
\bibitem{PhysRevLett.109.056805} A. C. Betz, F. Vialla, D. Brunel, C. Voisin, M. Picher, A. Cavanna, A. Madouri, G. Fe\`ve, J.-M. Berroir, B. Placais, and E. Pallecchi, Phys. Rev. Lett. {\bf 109}, 056805 (2012).
\bibitem{SciRep.3.3533} Q. Han, T. Gao, R. Zhang, Y. Chen, J. Chen, G. Liu, Y. Zhang, Z. Liu, X. Wu, and D. Yu, Sci. Rep. {\bf 3}, 3533 (2013).
\bibitem{JPCM.14.R501} J. J. Lin and J. P. Bird, Journal of Physics: Condensed Matter {\bf 14}, R501 (2002).
\bibitem{NanoLett.13.4305} R. Somphonsane, H. Ramamoorthy, G. Bohra, G. He, D. K. Ferry, Y. Ochiai, N. Aoki, and J. P. Bird, Nano Lett. {\bf 13}, 4305 (2013).

\bibitem{SciRep.7.10317} R. Somphonsane, H. Ramamoorthy, G. He, J. Nathawat, C.- P. Kwan, N. Arabchigavkani, Y.-H. Lee, J. Fransson, and J. P. Bird, Sci. Rep. {\bf 7}, 10317 (2017).
\bibitem{RevModPhys.81.109} A. H. Castro Neto, F. Guinea, N. M. R. Peres, K. S. Novoselov, and A. K. Geim, Rev. Mod. Phys. {\bf 81}, 109 (2009).
\bibitem{ApplPhysLett.101.093110} G. Bohra, R. Somphonsane, N. Aoki, Y. Ochiai, D. K. Ferry, and J. P. Bird, Appl. Phys. Lett. {\bf 101}, 093110 (2012).
\bibitem{PhysRevB.86.161405} G. Bohra, R. Somphonsane, N. Aoki, Y. Ochiai, R. Akis, D. K. Ferry, and J. P. Bird, Phys. Rev. B {\bf 86}, 161405(R) (2012).
\bibitem{QuantumTheorySolids} B.L. Altshuler, A.G. Aronov, D.E. Khmelnitskii, and A.I. Larkin, in Quantum Theory of Solids, edited by I.M. Lifshits (Mir, Moscow, 1982).

\bibitem{PhysRep.140.193} S. Chakravarty and A. Schmid, Phys. Rep. {\bf 140}, 193 (1986).
\bibitem{SolStComm.86.195} N. H. Shon and A. Ferraz, Sol. State Comm. {\bf 86}, 195 (1993).
\bibitem{QuantumKinetics} H. Haug and A. -P. Jauho, in Quantum Kinetics in Transport and Optics of Semiconductors (Springer-Verlag, Berlin/Heidelberg/New York, 1998).
\bibitem{JPhysSocJpn.67.2421} N. H. Shon and T. Ando, J. Phys. Soc. Jpn. {\bf 67}, 2421 (1998).
\bibitem{PhysRevLett.89.266603} H. Suzuura and T. Ando, Phys. Rev. Lett. {\bf 89}, 266603 (2002).

\bibitem{PhysRevB.73.125411} N. M. R. Peres, F. Guinea, and A. H. Castro Neto, Phys. Rev. B {\bf 73}, 125411 (2006).
\bibitem{PhysRevB.33.8216} K. K. Choi, D. C. Tsui, and S. C. Palmateer, Phys. Rev. B {\bf 33}, 8216 (1986).
\bibitem{PhysRevB.35.1039} P. A. Lee, A. D. Stone, and H. Fukuyama, Phys. Rev. B {\bf 35}, 1039 (1987).
\bibitem{NatureComms.6.7702} V. Tayari, N. Hemsworth, I. Fakih, A. Favron, E. Gaufres, G. Gervais, R. Martel, and T. Szkopek, Nature Comms. {\bf 6}, 7702 (2015).
\bibitem{PhysRevB.94.245404} N. Hemsworth, V. Tayari, F. Telesio, S. Xiang, S. Roddaro, M. Caporali, A. Ienco, M. Serrano-Ruiz, M. Peruzzini, G. Gervais, T. Szkopek, and S. Heun, Phys. Rev. B {\bf 94}, 245404 (2016).
\bibitem{PRL.108.036805} M. Liu, J. Zhang, C.-Z. Chang, Z. Zhang, X. Feng, K. Li, K. He, L.-L. Wang, X. Chen, X. Dai, Z. Fang, Q.-K. Xue, X. Ma, and Y. Wang, Phys. Rev. Lett. {\bf 108}, 036805 (2012).
\bibitem{NanoLett.13.48} M. Lang, L. He, X. Kou, P. Upadhyaya, Y. Fan, H. Chu, Y. Jiang, J. H. Bardarson, W. Jiang, E. S. Choi, Y. Wang, N.-C. Yeh, J. Moore, and K. L. Wang, Nano Lett. {\bf 13}, 48 (2013).

\end{thebibliography}
\end{document}